\documentstyle[11pt]{article}

\def\thebibliography#1{\list
 {$^{\arabic{enumi}}$}{\settowidth\labelwidth{[#1]}\leftmargin\labelwidth
 \advance\leftmargin\labelsep
 \usecounter{enumi}}
 \def\newblock{\hskip .11em plus .33em minus .07em}
 \sloppy\clubpenalty4000\widowpenalty4000
 \sfcode`\.=1000\relax}

\begin{document}

\makeatletter
\renewcommand{\@cite}[2]{{\footnotesize $^{#1\if@tempswa , #2\fi}$}}
\def\ncite#1{{\def\@cite##1##2{##1}\cite{#1}}}
\makeatother

\baselineskip=.7cm

{\large \noindent {\bf q DEFORMATION BY INTERTWINNING
\vspace{3mm}\noindent WITH APPLICATION TO THE SINGULAR OSCILLATOR}
\vspace{5mm}
\begin{quotation}
\noindent HARET C. ROSU$^{\dagger}$
\footnote{Electronic mail address: rosu@ifug3.ugto.mx}
and CARLOS CASTRO$^{*}$
\footnote{Electronic mail address: castro@hubble.cau.edu}

\noindent $^{\dagger}$
{\it Instituto de F\'{\i}sica, Universidad de Guanajuato, Apdo Postal 
E-143, Le\'on, Guanajuato 37150 Mexico

\noindent $^{*}$ {\it Center for Theoretical Studies of Physical Systems,
Clark Atlanta University, Atlanta, Georgia 30314 USA}

}

\vspace{5mm}
\noindent 
{\bf Abstract}.
We present a version of $q$-deformed calculus
based on deformed
counterparts of Darboux intertwining operators. The case in which
the deformed transformation function is of the vacuum type is detailed, but
the extension to counterparts of excited states used as Darboux transformation
functions is also formally discussed.
The method leads to second-order Fokker-Planck-like
deformed operators which may be considered as supersymmetric partners,
though for a sort of $q$-deformed open systems, i.e.,
those possessing $q$ nonlocal
drift terms, potential part, as well as $q$-spreaded vacuum fluctuations. The
undeformed limit
corresponds to the conservative case, since all $q$ nonlocalities
wash out. The procedure is applied to the $x^{-2}$ singular oscillator, for
which we also present a formal $q$ generalization of the Bagrov-Samsonov
coherent states.

\bigskip
Accepted at Phys. Lett. A on Nov. 29, 1999 [quant-ph/9808021 v4].

\end{quotation}

\begin{flushleft}
{\bf 1. INTRODUCTION}   
\end{flushleft}
\hspace{5mm}

At the present time, the $q$-deformed calculus is widely applied to many
physical problems \cite{Jac}, but although there are many advances
in mathematics, the attention of physicists is still
mainly captured by the $q$ harmonic oscillators
introduced by Arik and Coon, Biedenharn, and
Macfarlane \cite{pion}.
In this work, we present a general $q$-deformed
framework that next we apply to the singular oscillator, which is based on the
idea of using, as fundamental tools, some deformed counterparts of the
intertwining operators encountered in the area of Darboux transformations.
We have been motivated to write this paper by
recent results of Bagrov and Samsonov \cite{BS}, who provided an important
extension of the Darboux method by showing
that the application of the Darboux transformation/intertwining operator to
the coherent states of the initial quantum system
gives a set of states which may be treated as coherent states of the 
transformed system. 
Moreover, employing the example of the singular oscillator,
Samsonov \cite{S1} has
shown that the Darboux transformation for this case translates into the
transformation of the K\"ahler potential corresponding to a manifold of
varying curvature. This leads to a calculable distortion of the initial
phase space.

Our basic idea is to use the factorization property of deformed analogues
of the intertwining operators to build second-order deformed operators
which are of the Fokker-Planck type, containing
$q$ drift ($q$ first derivative) sectors, as well as contributions that,
when going to particular quantum mechanical problems,
may be shown to be modifications of the potential and zero point
sectors. In particular, we obtain a sort of $q$ nonlocal
``spatially-spreaded" zero point sector. In the undeformed
limit $q\rightarrow 1$, the $q$ drift sectors disappear, while the other
sectors get the conventional forms of the original Hamiltonian system
as we show for the singular oscillator.

The organization of the paper is the following. In the next section we present
our $q$-deformed intertwining method in general terms.
Then, in section 3, the theory of singular oscillator is briefly reviewed,
together with the recently introduced Darboux coherent states. In section 4,
the $q$-deformed intertwining method is applied to the singular oscillator,
introducing $q$-deformed Darboux coherent states as well, and
we end up with the conclusion.

\vspace{5mm}
{\bf 2. q DEFORMATION BY INTERTWINING} 

\vspace{5mm}
Here we present the general framework of the $q$-deformed calculus
based on deformed counterparts of the intertwining operators, which in
standard supersymmetric quantum mechanics are also known as
supersymmetric charges or supercharges.
When the method is applied to particular cases,
such as the singular oscillator (see section 4), one can get more
insight on the features of the procedure.

The independent variable is still
the real commutative half line, and therefore, from this point of view, our
version of the deformed calculus is similar to that previously used
by some authors who deformed the Coulomb problem \cite{coul}.
We shall employ symmetric definitions of the
$q$ number $[x]_q=\frac{q^{x}-q^{-x}}{q-q^{-1}}$
and $q$ derivative
\begin{equation}
D_{q}f(x)=\frac{f(qx)-f(q^{-1}x)}{x(q-q^{-1})}~.
\end{equation}
Some of the basic rules of Jackson's calculus \cite{Jac}, such as
$D_{q} x^{n}=[n]_{q}x^{n-1}$, $D_{q}^{2}x^n=[n]_q[n-1]_qx^{n-2}$,
$D_{q}(FG)=(D_{q}F)G(qx)+F(q^{-1}x)(D_{q}G)$
have been used in the calculations.
The definition of the $q$ exponential is
\begin{equation}
e_{q}(x)= \sum _{n=0}^{\infty}\frac{x^{n}}{[n]_{q}!}~,
\end{equation}
which reduces to the usual exponential function as $q\rightarrow 1$, and
is invariant under $q\rightarrow q^{-1}$.

As well known, in supersymmetric quantum mechanics a ground state (or vacuum)
can be written as $u_{0}(x)\propto e^{\int ^{x} W}$, where $W$ is a solution of
the corresponding Riccati equation usually called superpotential.
The main idea for the following is to consider a deformed ground 
state $u_{0q}$, in the sense
of deforming the exponential as in (2), and exploit the factorization
property of first-order deformed operators (supercharges) of the form
\begin{equation}
T_{01}^{q}=D_{q}-\frac{D_{q}u_{0q}}{u_{0q}}
\end{equation}
and
\begin{equation}
T_{02}^{q}=-D_{q}-\frac{D_{q}u_{0q}}{u_{0q}}~.
\end{equation}
These operators have been written by analogy to the
continuous intertwining operators. Thus, by employing the factorization
procedure, we would like to see what sort of second-order deformed operators
we shall obtain from the products $T_{02}^{q}T_{01}^{q}$ and
$T_{01}^{q}T_{02}^{q}$, respectively.

After a straightforward calculation we get the
following operators
\begin{equation}
\pounds _{0i}^{q}\equiv T_{02}^{q}T_{01}^{q}=
-D_{q}^{2}-
\Bigg[\left(\frac{D_{q}u_{0q}}{u_{0q}}\right)_{(x)}
-\left(\frac{D_{q}u_{0q}}{u_{0q}}\right)_{(x/q)}\Bigg]
D_{q}+\Bigg[\left(\frac{D_{q}u_{0q}}{u_{0q}}\right)^{2}_{\rightarrow x}+
D_{q}\left(\frac{D_{q}u_{0q}}{u_{0q}}\right)^{2}_{\rightarrow qx}\Bigg]
\end{equation}
and
\begin{equation}
\pounds _{0f}^{q}\equiv T_{01}^{q}T_{02}^{q}=
-D_{q}^{2}+
\Bigg[\left(\frac{D_{q}u_{0q}}{u_{0q}}\right)_{(x)}
-\left(\frac{D_{q}u_{0q}}{u_{0q}}\right)_{(x/q)}\Bigg]
D_{q}+\Bigg[\left(\frac{D_{q}u_{0q}}{u_{0q}}\right)^{2}_{\rightarrow x}-
D_{q}\left(\frac{D_{q}u_{0q}}{u_{0q}}\right)^{2}_{\rightarrow qx}\Bigg]~.
\end{equation}

If one multiplies (5) and (6) by (-1) a quite natural interpretation of the
operators $-\pounds _{0i}^{q}$
and $-\pounds _{0f}^{q}$ is as $q$-deformed Fokker-Planck
operators. They may be considered
as supersymmetric partners describing a pair of open systems that are
connected in an algebraic manner. Along with
the $q$ kinetic $ D_{q}^{2}$
term, they contain a $D_{q}$ drift part,
and a nonoperatorial potential contribution.
The latter two sectors have been delimited by square brackets in
(5) and (6). The directional subindices indicate the argument of the
solution to which the nonoperatorial terms are to be attached, whereas the
subindices in the drift coefficients show explicitly their arguments.
The main feature of the two operators is the $q$ nonlocality.
Using the Fokker-Planck interpretation, one may say that the
coefficient of the first derivative term is the $q$ derivative
of the drift potential.

The procedure can be extended easily to models for which the deformed Darboux
transformation functions correspond to excited
states of the form
\begin{equation}   \label{def1}
\psi _{nq}\propto P_{n}(x^2;q)e_{q}(-\beta x^2)~,
\end{equation}
where $P_{n}(x^2;q)$ is a deformed
counterpart of the special polynomial $P_{n}(x^2)$.
As a matter of fact, a third monomial factor is also allowed in (7).
In order to avoid any singularities in the undeformed limit,
it is convenient to perform an $i$-rotation
$x\rightarrow ix$, leading to
\begin{equation}      \label{u}
u_{pq}(x)\propto P_{pq}\left( y\right)
\exp _{q} \left( \beta x^2\right) ,\quad
y=-x^2/2 ,
\quad p=0,1,2,\ldots
\end{equation}
and use $u_{pq}(x)$ as the Darboux transformation function since
$P_{pq}(y)\ne 0$ $\forall x\ne 0$ when
$y=-x^2/2$$(<0)$ and, hence, the function $u_{pq}(x)$ is nodeless for $x>0$.
The intertwining operators can be calculated according to
$T_{p1}^{q}=D_{q}-\frac{D_q u_{pq}}{u_{pq}}$ and
$T_{p2}^{q}=-D_{q}-\frac{D_q u_{pq}}{u_{pq}}$, and again by exploiting the
factorization property one is led to second-order deformed operators of a
more complex structure than (5) and (6). Formally, one should merely
replace the subindex $0$ with $p$ in (5) and (6).
We shall apply this extension to the singular oscillator that belongs
to the class of models possessing excited states for which the $i$ rotation
works.

\vspace{5mm}
{\bf 3. THE SINGULAR OSCILLATOR}\\ 

\vspace{3mm}

The $x^{-2}$ singular quantum oscillator
appears in many physical problems, such as diatomic and polyatomic
molecules \cite{molec},
the Calogero problem \cite{Cal}, fractional statistics and anyons \cite{LM},
two-dimensional QCD \cite{MP}, and others.
Very recently, Dodonov, Man'ko and Rosa \cite{dmr}, following
Combescure's suggestion \cite{comb}, presented a detailed study
of the more general time-dependent quantum singular oscillator
as a model of a two-ion trap.
Especially in the latter case, various coherent states are an
important topic, because their mesoscopic superpositions commonly known as
Schr\"odinger cat-like states can be studied experimentally,
although those of Darboux type \cite{dcoh}
have not yet been investigated in the physics of traps. Moreover, in the
same context, discrete irregular effects of both
(quasi)periodic $\delta$-kick type and other discrete spatial type may prove
important since, as remarked in \cite{dmr}, the singular oscillator system
lies at the border between
linearity and nonlinearity, and chaotic behavior is a characteristic
feature of the overwhelming majority of deterministic nonlinear dynamical
systems \cite{dchaos}.

We now briefly sketch the approach of
Bagrov and Samsonov for a Darboux class of coherent states \cite{BS}.
The Hamiltonian of the (half-line/radial) quantum oscillator
with $x^{-2}$ singular part
\begin{equation}
h_0=-d^2/dx^2+x^2/4+b/x^2,\quad b >0,\quad x\in [0,\infty )
\label{1}
\end{equation}
has the $su(1,1)$ dynamical symmetry algebra and therefore is exactly
solvable. The more general time-dependent singular oscillator has been
already solved in 1971 \cite{71}.
In coordinate
representation the generators of this algebra are expressed in terms of the
harmonic oscillator annihilation $a$ and creation $a^{+}$ operators
\begin{equation}
\begin{array}{c}
K_0=\frac {1}{2}h_0,\quad K_{+}=\frac {1}{2}\left[
(a^{+})^2-b/x^2\right] ,\quad K_{-}=\frac {1}{2}\left[
a^2-b/x^2\right] ,\quad  \\ a=d/dx+x/2,\quad a^{+}=-d/dx+x/2~,
\label{2}
\end{array}
\end{equation}
and satisfy the standard commutation relations
\begin{equation}
\left[ K_0,K_{\pm }\right] =
\pm K_{\pm },\quad \left[ K_{-},K_{+}\right] =2K_0\ .
\end{equation}
For an irreducible representation the corresponding $su(1,1)$
Casimir operator takes the
constant value ${\cal C}=\frac 12\left[ K_{+}K_{-}+K_{-}K_{+}\right]
-K_0^2=3/16-b/4=k(1-k)$. The value of $k=1/2+(1/4)\sqrt{1+4b}$ defines the
ground state (vacuum) energy $E_0=2k$. The vacuum state $|0\rangle $ is
defined by the equations: $K_{-}\mid 0\rangle =0$ and $K_0\mid 0\rangle =k\mid
0\rangle $.
A very interesting feature of the ground state of the singular oscillator is
the presence of a centrifugal barrier of the type $b_{0}/x^2$, where
$b_{0}=(2k-3/2)(2k-1/2)$ that may be considered as produced by the zero
point fluctuations.

Other discrete Fock eigenfunctions $|n\rangle $ of the
Hamiltonian $h_0$ are defined by applying
successive powers of the raising operator $K_{+}$,
and in the coordinate representation they have the form
\begin{equation}
\psi _n(x)=\langle x\mid n\rangle =\left[ n!2^{1-2k}\Gamma
^{-1}(n+2k)\right] ^{1/2}x^{2k-1/2}L_n^{2k-1}(x^2/2)\exp (-x^2/4),
\end{equation}
where $L_n^{2k-1} (x^2/2)$ are the associated Laguerre polynomials.

Passing now to the $su(1,1)$ coherent states, we first mention their
rich variety in the literature \cite{scoh}, although one may consider as the
most appropriate those
introduced by Barut and Girardello \cite{BG} as the eigenstates of the
Weyl lowering operator $K_{-}$ of the $su(1,1)$ algebra.
In the approach of Bagrov and Samsonov the coherent states are the
kets
$
\mid z\rangle =
N_{0z}\sum_{n=0}^\infty
a_nz^n\mid n\rangle ,
$
where
\begin{equation}
N_{0z}^{2}=(1-z\overline{z})^{2k},\quad a_n=
(-1)^n\sqrt{\frac{\Gamma (2k+n)}{%
n!\Gamma (2k)}},\quad \left| z\right| <1.
\end{equation}

The linear manifold spanned by the coherent vectors $\{\mid z\rangle \}$
forms an
everywhere dense and overcomplete set in the Hilbert space ${\cal H}_{1}$
with the identity decomposition
$
\int _{\left| z\right| <1}\mid z\rangle \langle z\mid d\mu
(z)=1$, of
measure $ d\mu (z)=\frac{2k-1}\pi (1-z\overline{z})^{-2}dzd\overline{z}.
$

The Fourier coefficients $c_n$ of a vector $\mid \psi \rangle \in
{\cal H}_{1}$ with
respect to the basis $\{\mid n\rangle \}$ define a holomorphic
representation $\psi (z)$ of the vector $\mid \psi \rangle $ in the space of
functions which are holomorphic in the unit disk
\begin{equation}
\langle \overline{z}\mid \psi \rangle =N_{0z}\psi \left( z\right) ,\quad
\psi \left( z\right) =\sum\limits_{n=0}^\infty a_nc_nz^n.
\end{equation}

Let ${\cal L}$ be the linear span of entire  analytic functions
$\psi(z)$ defined in the unit disk such that
$\int _{|z|<1}\mid \psi (z)\mid ^2(1-|z|^2)^{2k}d\mu (z)<\infty .$
If in ${\cal L}$ one defines the inner product
\begin{equation}
\langle \psi _1\left( z\right) \mid \psi _2\left( z\right) \rangle
\equiv \int _{|z|<1}e^{-f^{(0)}}\overline{\psi }_1\left( z\right)
\psi _2\left(z\right) d\mu \left( z\right) =\langle \psi _1
\mid \psi _2\rangle ,
\end{equation}
where $\langle \psi _1\mid \psi _2\rangle $ is the inner product in the
Hilbert space ${\cal H}_{1}$ then the completion of ${\cal L}$ with respect
to this inner product gives the Hilbert space ${\cal H}_{2}$. The
function $f^{\left(0\right) }=\ln \left| \langle 0\mid z\rangle
\right| ^{-2}=\ln (1-z\overline{z})^{-2k}$ is the K\"ahler potential in the
unit disk. The corresponding metric has the form
\begin{equation}
ds^2=g_{z\bar z}
dzd\overline{z},\quad g_{z\bar z}=\frac{\partial ^2f^{\left( 0\right) }}{%
\partial z\partial \overline{z}}
=\frac{2k}{\left( 1-z\overline{z}\right) ^2}~.
\end{equation}
The imaginary part of this metric defines a symplectic
2-form $\omega =-igdz\wedge d\overline{z}$ and consequently a Poisson
bracket $\{F_1,F_2\}$ of the functions $F_1$ and $F_2$ defined on the unit
disk. The unit disk has now all the requirements of a phase space of
constant Gauss curvature ${\cal K}^{(0)}=-\frac 2k$.


In order to get Darboux-transformed coherent states
one should use
the standard Darboux transformation operator
${T}=-t_{u}(x)+d/{dx}=-u^{\prime }(x)/u(x)+d/{dx}$,
where the prime denotes the derivative with respect to $x$.
When acting on the solutions $\psi _n(x)$ of the initial Schr\"odinger
equation $h_0\psi _n(x)=E_n\psi _n(x),\quad E_n=2n+2k$,
it transforms them into the solutions of another Schr\"odinger equation
$h_1\varphi _n(x)=E_n\varphi _n(x)$,
$\varphi _n(x)=N_n{T}\psi _n(x)$,
with the same eigenvalues $E_n$. The factor $N_n$ is introduced to ensure
that the states $\varphi _n$ are
normalized to unity. The new exactly solvable Hamiltonian has the form
$h_1=h_0+A(x)$, where the potential difference is of Darboux type
$A(x)=-2(\ln
u)^{\prime \prime }$. The function $u=u(x)$
is the Darboux transformation function,
being a solution of the initial Schr\"odinger equation
$
h_0u(x)=\alpha u(x)
$
with $\alpha \leq E_0$.
Samsonov treated the simpler case
when $\alpha <E_0$, (more exactly he used $\alpha =-2(k+p)$ and an analytic
continuation in order to work with a nodeless transformation function). Thus,
$u(x)\ne 0$ $\forall x\in
(0,\infty )$ and $1/u(x)$ is not a square integrable function in the interval
$[0,\infty )$. In this  case $u\notin {\cal H}_{1}$ and the
set $\{\mid \varphi _n\rangle \}$
constitutes a complete basis in the Hilbert space ${\cal H}_{1}$ provided
the initial system $\{\mid \psi _n\rangle \}$ is complete.
In terms of the supersymmetric quantum mechanics this case corresponds to a
broken supersymmetry.
The operator ${T}^{+}=-t_{u}(x)-d/dx$ performs the transformation in
the inverse
direction $\label{psin}|\psi _n\rangle =N_n {T}^{+}|\varphi _n\rangle$,
and together with ${T}$ participates in the factorizations
${T}^{+}{T}=h_0-\alpha ,\quad {T}{T}^{+}=h_1-\alpha \ .$
The ${T}$ and ${T}^{+}$ operators are
well defined $\forall \psi \in {\cal H}_{1}$ and are conjugated to each
other with respect to the inner product in the ${\cal H}_{1}$ space.

We can now define, following Bagrov and Samsonov \cite{BS},
the Darboux-transformed coherent states
$\varphi _z(x)=N_{1z} {T}\psi _z(x)=N\sum\limits_{n=0}^\infty
b_nz^n\varphi _n$, where
$N=N_{0z}N_{1z}/N_0,\quad b_n=a_nN_0/N_n$.
From the factorization properties of the $T$ operators
one can derive the value of the normalization constant
$N_{1z}^{2}=(1-z\overline{z})/(4k+2p-2pz\overline{z})$, which,
as shown by Samsonov \cite{S1}, is responsible
for the Gauss varying curvature of the phase space disk.

\vspace{5mm}
\begin{flushleft}
{\bf 4. q-DEFORMED INTERTWINING FOR THE SINGULAR OSCILLATOR}
\end{flushleft}

\vspace{5mm}

For the singular oscillator the deformed vacuum states are of the type
$u_{0q}\propto x^{\gamma}e_q(\beta x^2)$,
where $\gamma =2k-1/2$ and  $\beta =\pm 1/4$ for the irregular and regular
vacuum respectively. The irregular vacuum is the one not well-behaved
asymptotically, but, as known, can be used as Darboux transformation function.
The first-order deformed operators read
\begin{equation}
T_{01}^{q}=D_{q}-\frac{D_{q}u_{0q}}{u_{0q}}=
D_q-\frac{[\gamma]_{q}}{x}\epsilon _{q}(x^2)
-\beta _{q}(x^2) x~,
\end{equation}
\begin{equation}
T_{02}^{q}=-D_{q}-\frac{D_{q}u_{0q}}{u_{0q}}=
-D_q-\frac{[\gamma]_{q}}{x}\epsilon _{q}(x^2)
-\beta _{q}(x^2)x~,
\end{equation}
where
\begin{equation}
\epsilon _{q}(x^2)=\frac{e_{q}(\beta q^2x^2)}{e_{q}(\beta x^2)}~,
\end{equation}
\begin{equation}
\beta _{q}(x^2)=4\beta q^{-\gamma}\left(\frac{qe_{q}(q\beta x^{2})+
q^{-1}e_{q}(q^{-1}\beta x^{2})}{e_{q}(\beta x^2)}\right)~.
\end{equation}

The second-order $q$-deformed operators are given by
$$
\pounds _{0i}^{q}\equiv T_{02}^{q}T_{01}^{q}=
-D_{q}^{2}-
\Bigg[\left(\frac{[\gamma]_{q}}{x}(\Delta \epsilon _{q})+
(\Delta \beta _{q})x\right)D_{q}\Bigg]+
$$
$$
\Bigg[\left(\frac{[\gamma]_{q}^2}{x^2}\epsilon _{q}^{2}(x^2)
+\beta _{q}^{2}(x^2)x^2
\right)_{\rightarrow x}
-\left(\frac{[\gamma ]_{q}}{x^2}\epsilon _{q}(q^2x^2)
-\frac{q[\gamma]_{q}}{x}
(D_{q}\epsilon _{q}(x^2))
-q(D_{q}\beta _{q}(x^2))x\right)_{\rightarrow qx}\Bigg]
$$
\begin{equation}
+\Bigg[\left(2[\gamma]_{q}\beta _{q}
(x^2)\epsilon _{q}(x^2)\right)_{\rightarrow x}+
\left(\beta _{q}(q^{-2}x^2)\right)_{\rightarrow qx}\Bigg]
\end{equation}
and

$$
\pounds _{0f}^{q}\equiv T_{01}^{q}T_{02}^{q}=
-D_{q}^{2}+
\Bigg[\left(\frac{[\gamma]_{q}}{x}(\Delta \epsilon _{q})+
(\Delta \beta _{q})x\right)D_{q}\Bigg]+
$$
$$
\Bigg[\left(\frac{[\gamma]_{q}^2}{x^2}\epsilon _{q}^{2}(x^2)
+\beta _{q}^{2}(x^2)x^2
\right)_{\rightarrow x}
+\left(\frac{[\gamma ]_{q}}{x^2}\epsilon _{q}(q^2x^2)-\frac{q[\gamma]_{q}}{x}
(D_{q}\epsilon _{q}(x^2))
-q(D_{q}\beta _{q}(x^2))x\right)_{\rightarrow qx}\Bigg]
$$
\begin{equation}
+\Bigg[\left(2[\gamma]_{q}\beta _{q}(x^2)\epsilon _{q}
(x^2)\right)_{\rightarrow x}-
\left(\beta _{q}(q^{-2}x^2)\right)_{\rightarrow qx}\Bigg]~,
\end{equation}
where
\begin{equation}
\Delta \epsilon _{q}=\epsilon _{q}(x^2)- q\epsilon _{q}(q^{-2}x^2)~,
\end{equation}
and
\begin{equation}
\Delta \beta _{q}=\beta _{q}(x^2)- q^{-1}\beta _{q}(q^{-2}x^2)~.
\end{equation}

In (21) and (22) one can see a nonlocal potential part in the second square
bracket and a nonlocal zero point contribution in the third square bracket.
Moreover, in the $q\rightarrow 1$ limit,
the potential and zero point sectors take forms
identical to those of the undeformed singular oscillator. More precisely,
the undeformed limits read ($d=d/dx$)
\begin{equation}
\pounds _{0i}^{1}\equiv h_{0}=-d^2+\frac{\gamma(\gamma -1)}{x^2}+
\beta ^{2}x^2+\beta(2\gamma+1)
\end{equation}
and
\begin{equation}
\pounds _{0f}^{1}\equiv h_{1}=-d^2+\frac{\gamma(\gamma +1)}{x^2}+
\beta ^{2}x^2+\beta(2\gamma-1)~.
\end{equation}
Thus, when  $q\rightarrow 1$ one gets the conservative limit of the
open (driven) systems described by the operators $\pounds ^{q}$. Notice also
that the difference between the Hamiltonian partners (25) and (26)
can be assigned to a change of the balance between the centrifugal barrier
and the vacuum fluctuations.


In the more general case of the excited
states of the singular oscillator, the deformed counterparts can be written
in the form
\begin{equation}
\psi _{nq}\propto x^{\gamma}L_{n}^{\gamma -1/2}(x^2;q)e_{q}(-\frac{1}{4}x^2)~,
\end{equation}
where $L_{n}^{\gamma-1/2}(x^2;q)$ is a deformed
Laguerre polynomial (see \cite{KS}). 
One can see that we are in the class of models mentioned at the end of
section 2. Thus, in order to avoid any singularities,
it is convenient to perform an $i$-rotation
$x\rightarrow ix$, leading to
\begin{equation}
u_{pq}(x)\propto x^{\gamma}L_{pq}^{\gamma -1/2}\left( y\right)
\exp _{q} \left( x^2/4\right) ,\quad
y=-x^2/2 ,
\quad p=0,1,2,\ldots
\end{equation}
and use $u_{pq}(x)$ as the Darboux transformation function since
$L_{pq}^{\gamma -1/2}(y)\ne 0$ $\forall x\ne 0$ when
$y=-x^2/2$$(<0)$ and, hence, the function $u_{pq}(x)$ is nodeless for $x>0$.

The intertwining operators can be calculated according to
$T_{p1}^{q}=D_{q}-\frac{D_q u_{pq}}{u_{pq}}$ and
$T_{p2}^{q}=-D_{q}-\frac{D_q u_{pq}}{u_{pq}}$, and by employing the
factorization property one is led to second-order deformed operators of a
more complex structure than in the vacuum case.

Moreover, a whole manifold of Bagrov-Samsonov Darboux-transformed
wavefunctions corresponding to the deformed number states of the singular
oscillator can be introduced by applying various $T_{p1}^{q}$ operators
\begin{equation}
\phi _{np}^{q}(x)   
\propto T^{q}_{p1}\psi _{nq}(x)~.
\end{equation}
The corresponding q-deformed Darboux coherent states are defined by the
expansion 
\begin{equation}
\phi _{zp}^{q}(x)    
\propto\sum (b_{np}^{q}z^{n})\phi _{np}^{q}(x)~,
\end{equation}
and they fulfill the identity decomposition of the type
\begin{equation}
\int _{|z|<1}|\phi ^{q} _{zp}\rangle \langle\phi ^{q}_{zp}|d_{q}\nu(z) =1~,
\end{equation}
where the integral is of Jackson type, i.e., $\int _{0}^{a}f(x)d_{q}x=a
(q^{-1}-q)\sum _{n=0}^{\infty}q^{2n+1}f(q^{2n+1}a)$.

The calculation of the Darboux distortion of the dissipative phase space
of commutative $q$-deformed systems is beyond the scope of this work.
Nevertheless, we point here that
while in the conservative case, because of the isospectrality property,
the Darboux distortion is consistent with the change of the Hamilton
function in such a manner that the curves of a constant energy
(i.e., the classical trajectories) in the Darboux-transformed phase space
remain
unchanged \cite{S1}, the deformed case corresponds to driven and dissipative
systems and will miss this property.

\vspace{5mm}
\begin{flushleft}
{\bf 5. CONCLUSION}
\end{flushleft}

\vspace{3mm}

In conclusion, we presented here a new approach to a commutative
$q$-deformed
calculus, which is based on a special type of $q$-deformed
intertwining operators.
As an example, we applied our procedure to the important solvable case of
the $x^2 +x^{-2}$ singular oscillator.
Mesoscopic physics is the realm of open systems, being dominated by nonlocal
and decoherence effects.
We have shown by Darboux methods that the $q$ deformation is a different
means to introduce in the dynamics drifts and dissipation and possibly
spatial discreteness (e.g., by discretizing the deformation
parameter \cite{plf}), all endowed with $q$ nonlocal features.

\vspace{0.5cm}
\noindent
This work has been partially supported by CONACYT project 458100-5-25844E.


\end{document}